\documentclass[fleqn,usenatbib]{mnras}

\usepackage{newtxtext,newtxmath}
\usepackage{longtable}

\usepackage[T1]{fontenc}
\usepackage{ae,aecompl}

\usepackage{graphicx}	
\usepackage{amsmath}	

\title[Average solar flare time profile]{The morphology of average solar flare time profiles from observations of the Sun's lower atmosphere}

\author[Kashapova et al.]{Larisa K. Kashapova,$^{1}$
Anne-Marie Broomhall$^{2}$\thanks{E-mail: A-M.Broomhall@warwick.ac.uk},
Alena I. Larionova,$^{3}$
\newauthor
Elena G. Kupriyanova$^{3}$, and Ilya D. Motyk$^{4}$
\\
$^{1}$Institute of Solar-Terrestrial Physics, SB Russian Academy of Sciences, Lermontov str. 126a, 664033, Irkutsk, Russia\\
$^{2}$Centre for Fusion, Space and Astrophysics, Department of Physics, University of Warwick, UK \\
$^{3}$Department of Radio Astronomical Research, Central Astronomical Observatory at Pulkovo of the RAS, \\ Pulkovskoe Shosse 65/1, Saint Petersburg 196158, Russian Federation \\
$^{4}$ Faculty of Physics, Irkutsk State University, 20 Gagarin Blvd., Irkutsk, 664003, Russia
}

\date{Accepted XXX. Received YYY; in original form ZZZ}

\pubyear{2020}

\begin{document}
\label{firstpage}
\pagerange{\pageref{firstpage}--\pageref{lastpage}}
\maketitle

\begin{abstract}
We study the decay phase of solar flares in several spectral bands using a method based on that successfully applied to white light flares observed on an M4 dwarf. We selected and processed 102 events detected in the Sun-as-a-star flux obtained with SDO/AIA images in the 1600~{\AA} and 304~{\AA} channels and 54 events detected in the 1700~{\AA} channel. The main criterion for the selection of time profiles was a slow, continuous flux decay without significant new bursts. The obtained averaged time profiles were fitted with analytical templates, using different time intervals, that consisted of a combination of two independent exponents or a broken power law. The average flare profile observed in the 1700~{\AA} channel decayed more slowly than the average flare profile observed on the M4 dwarf. As the 1700~{\AA} emission is associated with a similar temperature to that usually ascribed to M dwarf flares, this implies that the M dwarf flare emission comes from a more dense layer than solar flare emission in the 1700~{\AA} band. 
The cooling processes in solar flares were best described by the two exponents model, fitted over the intervals t1=[0, 0.5]$t_{1/2}$ and t2=[3, 10]$t_{1/2}$ where $t_{1/2}$ is time taken for the profile to decay to half the maximum value. The broken power law model provided a good fit to the first decay phase, as it was able to account for the impact of chromospheric plasma evaporation, but it did not successfully fit the second decay phase.

\end{abstract}




\section{Introduction}

Flares are explosive events that occur in solar and stellar atmospheres. Flares are observed across a wide range of wavelengths, such as radio, visible, X-rays and gamma rays and the emission responsible for these observations originates from many different regions within solar and stellar atmospheres, from photospheres to coronae. It is generally believed that both solar and stellar flares are produced by the same mechanism, namely magnetic reconnection \citep{2011LRSP....8....6S}. However, discrepancies are readily observed, most notably the energy of the flares, which, for stellar flares, can often be several orders of magnitudes larger than even the largest solar flares.

While stellar flares have been studied for decades, it is only recently, through observations made by NASA's \textit{Kepler} mission \citep{2010Sci...327..977B}, that a statistically large sample of flares for a single star has been obtained. \citet{Davenport2014ApJ} detected over 6,000 flares on an M dwarf star in just 11 months of \textit{Kepler} data. Using a subset, containing 885 flares, which were classified as ``classical'' because they contained just a single peak, \citeauthor{Davenport2014ApJ} created an empirical flare template, that was well-represented by a polynomial  in the fast rising phase, and two exponential decays, one each for the impulsive and gradual decay phases respectively. The first decay phase corresponds to radiative cooling losses and the second one is thought to be related to thermal conduction losses \citep{2004psci.book.....A}. Considering the large number of flares used, the scatter around this flare template is remarkably low. 

\textit{Kepler} obtained broadband white light observations (between 4200 and 9000~{\AA}) of stellar intensity and has vastly increased the number of solar-like stars on which flares have been observed, prompting a number of statistical studies \citep[e.g.][]{2015EP&S...67...59M, 2019ApJ...876...58N}, many of which consider flaring rates, with a particular interest in determining how likely the Sun is to produce an energetic superflare. While flares are a common feature of \textit{Kepler} light curves, the same cannot be said for white light flare observations of the Sun. Solar white light flares are rare because they tend to be relatively short in duration (a few minutes) and have a low contrast, making ``Sun-as-a-star'' observations of white light flares particularly challenging. Nevertheless, \citet{2011A&A...530A..84K} performed a statistical study of Sun-as-a-star flares observed in a number of data sets, spanning a range of wavelengths, including Total Solar Irradiance (TSI). \citeauthor{2011A&A...530A..84K} demonstrated that white light emission is ubiquitous in solar flares, regardless of the energy of that flare, and that, while there is some dependence on flare strength, the blackbody temperature of solar flares is between $8,000$--$10,000$\,K, which is similar to estimates for M dwarf flare temperatures \citep{1992ApJS...78..565H}. In a similar analysis to that of \citeauthor{Davenport2014ApJ}, \citeauthor{2011A&A...530A..84K} produced average flare profiles, using 2,100 flares, and found that the TSI profile only contained the impulsive phase, with no evidence of the gradual decay phase. \citet{2011A&A...530A..84K} speculated that this could be because the gradual decay was below the noise level. 

However, using resolved observations of the Sun, it has been shown that white light flares are common even for relatively weak flares \citep[e.g.][]{2003A&A...409.1107M, 2006SoPh..234...79H, 2008ApJ...688L.119J}. \citet{2016ApJ...833...50K} use highly resolved white light observations of a single flare to produce a large number of light curves within the flaring region. The authors find that 58\% of the light curves can be well represented by a single-component decay phase, while the other 42\% are better represented by two components, akin to the morphology found by \citet{Davenport2014ApJ}. \citeauthor{2016ApJ...833...50K} also found that, where two phases are favoured, the cooling times correspond to those expected for the chromosphere and corona respectively \citep{2006ApJ...641.1210X}. 

\citet{2017ApJ...851...91N} used the \textit{Helioseismic and Magnetic Imager} (HMI) onboard the \textit{Solar Dynamics Observatory} (SDO) to produce white light flare light curves, finding that, for a given energy, solar white light flares are an order of magnitude longer than stellar superflares observed with \textit{Kepler}. Although, it is worth noting that the energies of the solar and stellar flares do not overlap, and so this result is based on extrapolations. Furthermore, the solar light curves were obtained by selecting spatial regions corresponding to the flare, rather than using Sun-as-a-star data. Although potential explanations for the discrepancy are proposed in terms of differences in cooling or magnetic field strength, the authors remain understandably cautious, stating that a better understanding of the mechanisms responsible for white light emission would help clarify the situation.

The exact emission mechanisms responsible for white light flares are not yet totally clear \citep[see discussions in][and references therein]{2016ApJ...816...88K, 2016ApJ...833...50K}. It is generally assumed that continuum white light emission originates from the region encompassed by the mid-photosphere and lower chromosphere. White light flares can show good temporal correspondence with hard X-ray emission \citep[e.g.][]{1995A&AS..110...99F}, implying that energetic electrons play an important role. While these electrons are capable of reaching the chromosphere, the mechanism by which the energy is transported to the photosphere is still debated. To determine how analogous solar flare morphology is to the white light stellar flares observed by \citet{Davenport2014ApJ} this article considers $304$~{\AA}, $1600$~{\AA} and $1700$~{\AA} data from the \textit{Atmospheric Image Assembly} \citep[AIA;][]{Lemen2012SoPh_AIA} instrument onboard the SDO. Although these wavelengths are not traditionally considered as white light, and are outside the waveband observed by \textit{Kepler}, the lines are used here because they originate from various locations within the photosphere and chromosphere (as described in Section \ref{sec:Data}).

Although TSI and Sun-as-a-star SDO/HMI data may be more akin to Kepler data, solar flares are infrequently observed in these data, even in the case of large eruptive events. As a result, analyses often rely on assumed theoretical models. For example, \cite{Emslie2012ApJ} had to complement the direct measurements of bolometric irradiance by estimations based on modelling.  Moreover, the temporal resolution of SDO/AIA data is less than 1 minute, in contrast to TSI. This condition is important for the analysis of time profiles of solar flares which are more dynamic than stellar ones.

A better characterisation of the underlying shape of a flare permits studies of more transient flare light curve features, such as quasi-periodic pulsations \citep[QPPs; see][for recent reviews]{2016SoPh..291.3143V, 2020STP.....6a...3K}, which are notoriously difficult to detect and characterise robustly. Detection mechanisms that rely on detrending or model fitting \citep[e.g.][]{2018SoPh..293...61D, 2017A&A...600A..78P, 2019ApJS..244...44B} would benefit from information concerning the underlying flare shape.

This article aims to investigate the solar-stellar flare connection, and emission mechanisms associated with white light flares by comparing the flare profile associated with various layers of the Sun's lower atmosphere with the flare profile associated with white light observations of flares on an M dwarf obtained by \citet{Davenport2014ApJ}. The target of the study was to provide an instrument for the analysis of cooling during the decay phase and revealing cases related to additional sources of energy release. As previously mentioned, we use SDO/AIA data from the 1600~{\AA}, 1700~{\AA} and 304~{\AA} channels. In Section \ref{sec:Data}, we describe the data in more detail, including how they were combined to determine median flare profiles for each channel. These median flare profiles are then fitted with both a combination of two exponential functions and a broken power law, as described in Section \ref{sec:fitting}. The fits are discussed in detailed Section \ref{sec:discussion}, and the main conclusions are summarised in Section \ref{sec:conclusions}.


\section{Observations and data processing}
\label{sec:Data} 
We obtained the total flux of the Sun-as-a-star using the images obtained  by AIA \citep{Lemen2012SoPh_AIA}  on-board SDO. The channels used here demonstrate the chromospheric and photospheric emission \citep{ODwyer2010A&A}: 1600~{\AA} (transition region and upper photosphere), 1700~{\AA} (temperature minimum, photosphere) and 304~{\AA} (chromosphere, transition region). The image cadence is 24~s for the 1600~{\AA} and 1700~{\AA} channels and 12~s for the 304~{\AA} channel.

The initial flare selection was performed using the GOES flare catalogue \footnote{\url{https://hesperia.gsfc.nasa.gov/goes/goes_event_listings/}}. Each event was visually inspected and was required to have a ``classical" flare time profile, consisting of a fast rise followed by a slow decay without any flattening or additional peaks in soft X-rays (SXR). When determining which flares to include in our sample the time interval onset was taken to be about 10 minutes before the flare maximum in GOES X-ray flux and the duration was defined as the length of time taken for the flux to decrease to the pre-flare level. More rigorous time scales were defined in the subsequent analysis, as described below. The initial  list consisted of 359 flares from B5 to X9.3 GOES class. 

After processing the AIA images with the standard package of SolarSoftware \citep[SSW]{1998ESASP.417..225B, 1998SoPh..182..497F}, we obtained the total flux of the whole image for each time moment and produced a preliminary time profile of the flare. Thus, we processed the AIA images as if they were Sun-as-a-star or without an extracting the flare area and the resultant flux was equivalent to data of instruments observing without spatial resolution.  To enable a combined analysis of the flares of different strength and duration, the profiles were normalised in both flux and time. First, each time profile was normalised using the flux maximum observed in each flare, meaning the maximum in normalised flux was unity for all flares in the sample. Second, we defined the time moment of the flux maximum as zero, so the time of the rise phase had negative values and the decay phase time had positive values. Finally, to get the same time scale for events of different duration, we presented the time series in normalised units defined by the time taken for the intensity to decrease to half the maximum for each time profile ($t_{1/2}$), following the methodology of \cite{Davenport2014ApJ}. Then the duration of each time profile was limited to a range of --5 to 10 time bins ($t_{1/2}$). The time profiles in each channel were again checked to ensure there were no additional peaks during the decay phase that were missed when selecting the initial sample. The criteria for inclusion in the final sample was that any fluctuations in the time profile should be less than 30$\%$ of the flux value. We used the time profile derivative for control of this criterion. Application of this criterion also minimises uncertainties in the determination of the $t_{1/2}$ value. As we processed solar flares from X to B GOES class and saturation in emission could not be excluded, this current criteria allowed us to escape significant contribution from saturation when forming time profiles.

\begin{figure}
	\includegraphics[width=\columnwidth]{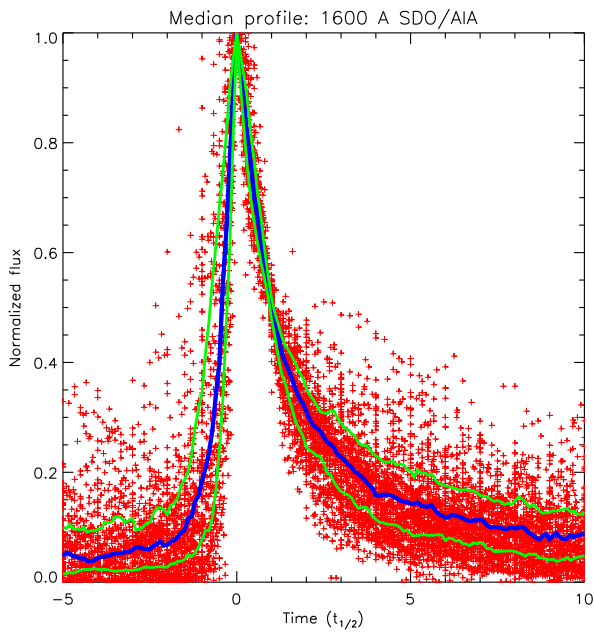}
    \caption{Time profile of solar flare emission at 1600~\AA. Red data points are the individual data from the 102 flares in the sample, thick blue line shows the median values and the thin green lines show the interquartile ranges.}
    \label{fig:1600_prof}
\end{figure} 

\begin{figure}
	\includegraphics[width=\columnwidth]{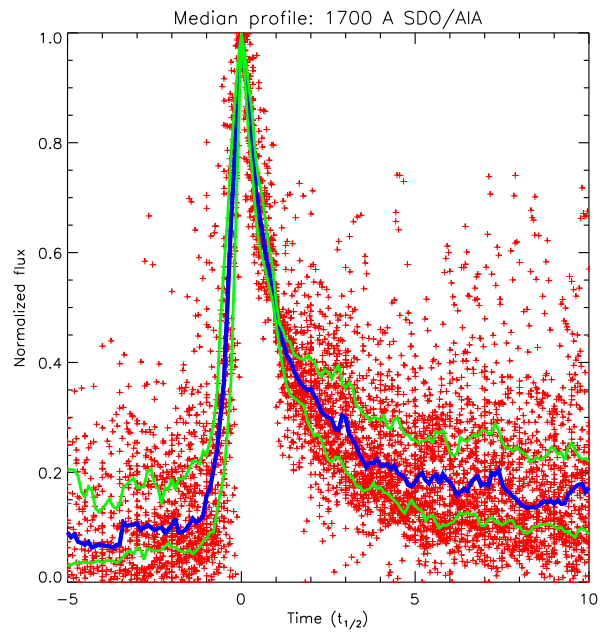}
    \caption{Time profile of solar flare emission at 1700 \AA. Line colours and data symbols are as described in Figure \ref{fig:1600_prof}, except only 54 flare profiles were included. }
    \label{fig:1700_prof}
\end{figure}

\begin{figure}
	\includegraphics[width=\columnwidth]{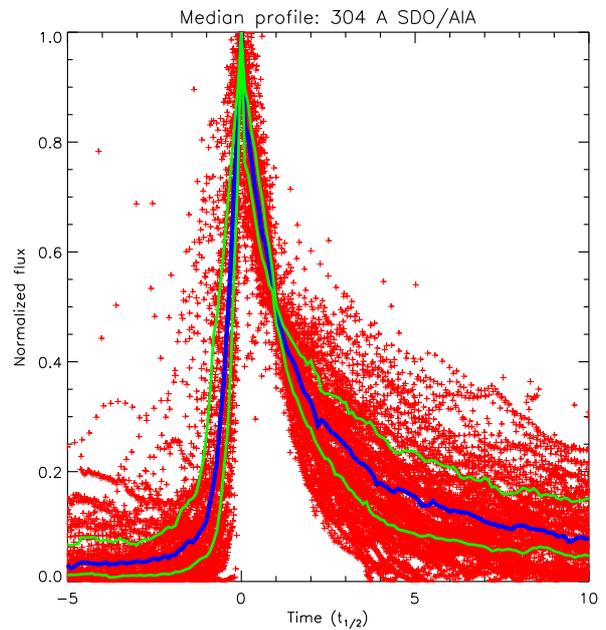}
    \caption{Time profile of solar flare emission at 304 \AA. Line colours and data symbols are as described in Figure \ref{fig:1600_prof}.}
    \label{fig:304_prof}
\end{figure}

The final sample contained 105 events in total from the 1600~{\AA} and 304~{\AA} channels (104 from 1600~{\AA} and 102 from 304~{\AA}). All profiles are presented on the same plot, with data shown as red crosses in Figures~\ref{fig:1600_prof} and \ref{fig:304_prof} for the 1600~{\AA} and 304~{\AA} channels, correspondingly. The X class flare ratio is 6$\%$, the M class flare ratio is 58$\%$, the C class flare ratio is 35$\%$ and B class flare ratio is 1$\%$. Only 53 time profiles of emission in the 1700~{\AA} channel showed both significant response and satisfied the all criteria outlined above (see Figure~\ref{fig:1700_prof}). 

To construct the average time profile we re-sampled time profiles to a time resolution of 0.001$t_{1/2}$ using linear interpolation, as was done by \cite{Davenport2014ApJ}. The average time profiles were defined as the median flux value computed at each time bin and are shown in Figures~\ref{fig:1600_prof}--\ref{fig:304_prof} as blue lines. One can see that the dispersion of initial data is significantly higher in the 1700~{\AA} band plot, despite containing a smaller number of flare profiles. 

\begin{figure}
	\includegraphics[width=\columnwidth]{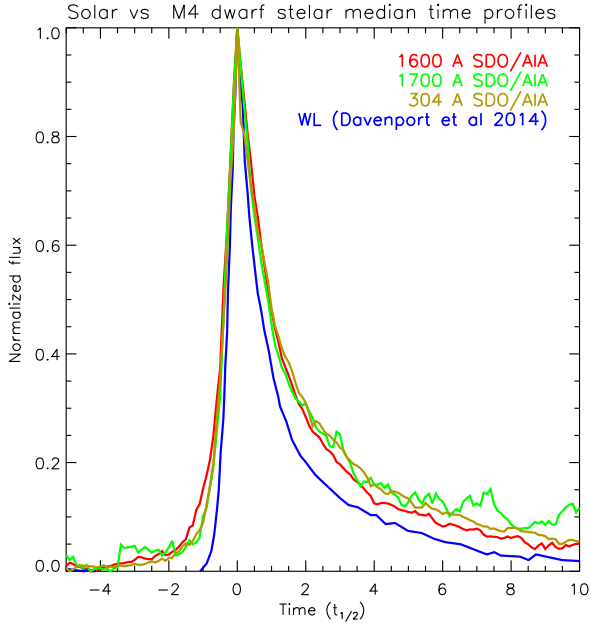}
    \caption{ Comparison of the obtained time profiles of solar flare emission with the time profile of the M4 dwarf.}
    \label{fig:1600_model}
\end{figure}

The standard errors, as shown in Figures~\ref{fig:1600_prof}--\ref{fig:304_prof}, are given by the median of values above and below the average value for high ($er_h$) and low values ($er_l$), correspondingly. This is also referred to as the interquartile range. As these errors are mainly asymmetric during the decay phase we used half of the sum of these two values ($\sigma =(er_h+er_l)/2$) as the error for the function fitting. The time-averaged time profiles obtained for the solar flares demonstrate the similar behaviour to the M4 dwarf flare time profile obtained by \cite{Davenport2014ApJ} (see in Figure~\ref{fig:1600_model}). During the decay phase, the evolution is fully agreed up to $t_{1/2}$ equal about 1 . After this moment, the solar intensity decays slower in all three wavelength bands relative to the M4 dwarf flare emission.

\section{Decay phase fitting} \label{sec:fitting}
\citet{Gryciuk2017SoPh} determined an analytic template of the decay phase of a solar flare composed of a single exponent with various coefficients taking into account peculiarities of the flare profile shape. The template was derived and applied to soft X-ray observations that describe coronal plasmas. However, the decay phase of a solar flare is a composition of the cooling process and various heating processes, especially in lower layers of the solar atmosphere. As our time profiles clearly demonstrated a steep initial decay followed by a slower decline in flux, we decided to apply the two-phase model. This model was successfully used by \cite{Davenport2014ApJ} for the flares observed on an M4 dwarf. Despite the difference in the atmospheric structure (temperature and density stratification) between the Sun and M dwarfs, the flare evolution is similar in both cases \citep{Allred_2006ApJ}.

\begin{figure}
	\includegraphics[width=\columnwidth]{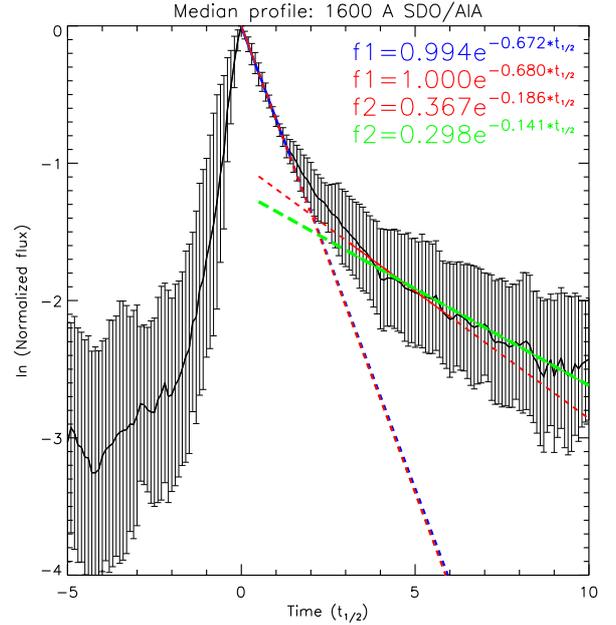}
    \caption{Average (median) time profile of solar flare emission at 1600~{\AA} (black with error bars) fitted by two exponential functions for different periods. Red, dashed lines and red text in the legend correspond to the t1=[0, 0.5]$t_{1/2}$ and t2=[3, 6]$t_{1/2}$. Blue dashed lines and blue text in the legend correspond to t1=[0, 1.5]$t_{1/2}$, and light green solid lines and light green text in the legend correspond to t2=[3, 10]$t_{1/2}$. }
    \label{fig:1600_fit}
\end{figure}
\begin{figure}
	\includegraphics[width=\columnwidth]{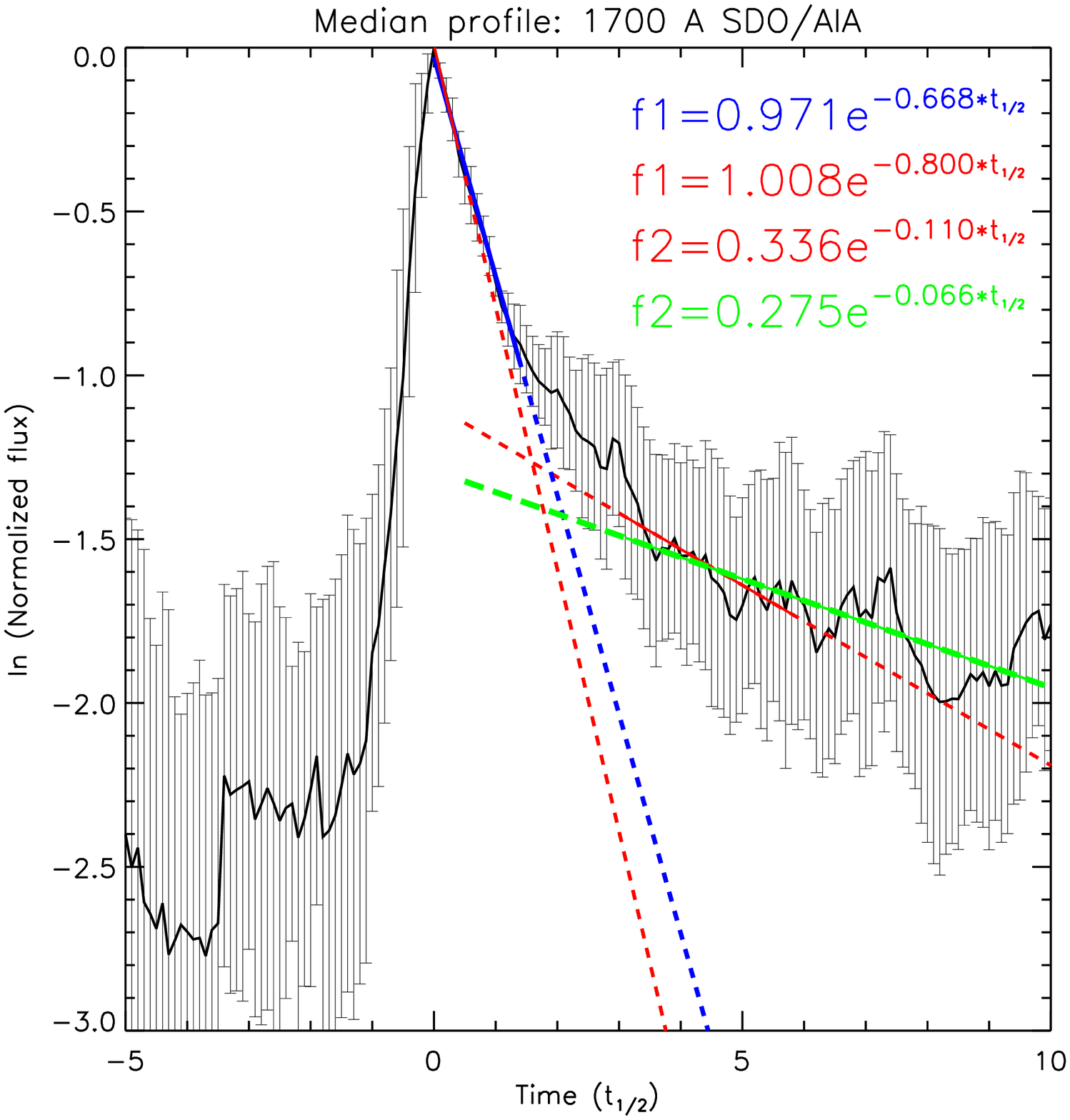}
    \caption{Average (median) time profile of solar flare emission at 1700~{\AA} fitted by two exponential functions for different periods. The line and text colours correspond to the same period as on Figure~\ref{fig:1600_fit}.
    }
    \label{fig:1700_fit}
\end{figure}
\begin{figure}
	\includegraphics[width=\columnwidth]{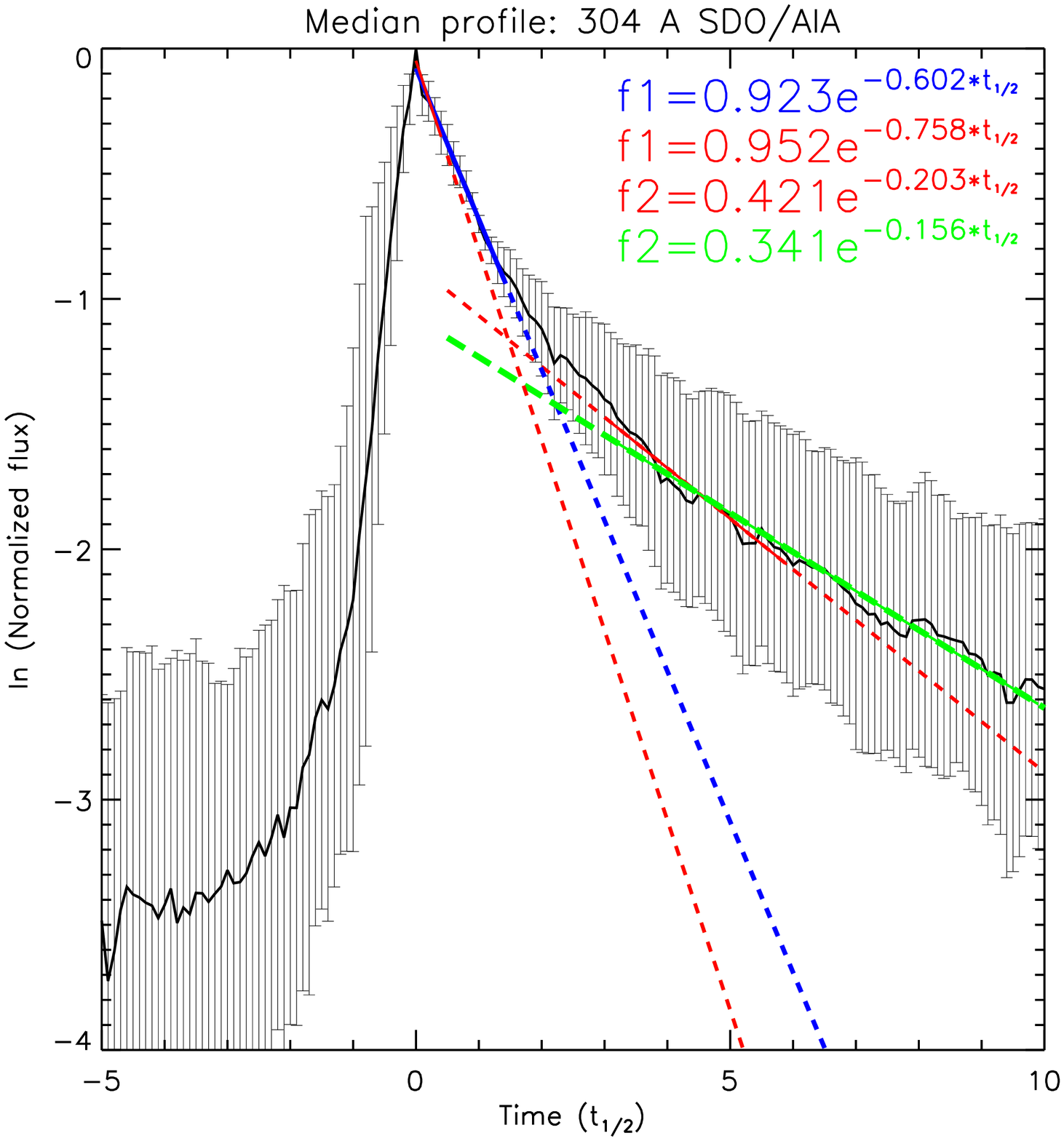}
    \caption{Average (median) time profile of solar flare emission at 304~{\AA} two exponential functions for different periods. The line and text colours correspond to the  same period as on Figure~\ref{fig:1600_fit}.}
    \label{fig:304_fit}
\end{figure}

The average time profiles, obtained using the natural logarithm of flux, are presented in Figures~\ref{fig:1600_fit}--\ref{fig:304_fit}. The flux data were fitted using two straight lines, $f_1(t)=a_1\exp^{b_1t_{1/2}}$ and $f_2(t)=a_2\exp^{b_2t_{1/2}}$, corresponding to the two different decay phases. Initially, the same time regions as \cite{Davenport2014ApJ} were used for the fitting. These ranges are $0 < t_{1/2}< 0.5$ for $f_1(t)$ and $3 < t_{1/2}< 6$ for $f_2(t)$. The results are shown in red in Figures~\ref{fig:1600_fit}--\ref{fig:304_fit}. In addition, alternative time regions were chosen to improve the fitting of each phase:  $0 < t_{1/2}< 1.5$ for $f_1(t)$ (shown in blue) and $3 < t_{1/2}< 10$ for $f_2(t)$ (shown in green). The results of the fitting are shown in Table~\ref{tab:table1}.

\begin{table}
	\centering
	\caption{Results of fitting by two exponential functions $f_1=a_1\exp^{-b_1t_{1/2}}$ and $f_2=a_2\exp^{-b_2t_{1/2}}$.}
	\label{tab:table1}
	\begin{tabular}{lcccc} 
		\hline
	Fit interval& &1600 \AA & 1700 \AA & 304 \AA \\
		\hline
	 $t_{1/2}$<0.5&$a_1$&1.000$\pm$0.001&1.008$\pm$0.010&0.952$\pm$0.034\\
	 $t_{1/2}$<1.5&$a_1$&0.994$\pm$0.006&0.971$\pm$0.013&0.923$\pm$0.013\\
	 
	 $t_{1/2}$<0.5&$b_1$&0.680$\pm$0.005&0.800$\pm$0.040& 0.758$\pm$0.147\\
	 $t_{1/2}$<1.5& $b_1$&0.672$\pm$0.007&0.668$\pm$0.016&0.601$\pm$0.018\\
	
	 3<$t_{1/2}$<6&$a_2$&0.367$\pm$0.016&0.336$\pm$0.021&0.421$\pm$0.018\\
	 3<$t_{1/2}$<10&$a_2$&0.298$\pm$0.007&0.275$\pm$0.011&0.341$\pm$0.007\\
	
	 3<$t_{1/2}$<6&$b_2$&0.186$\pm$0.010&0.110$\pm$0.013&0.203$\pm$0.009 \\ 3<$t_{1/2}$<10&$b_2$&0.141$\pm$0.004&0.066$\pm$0.006&0.156$\pm$0.003\\
	
		\hline
	\end{tabular}
\end{table}

\begin{figure}
	\includegraphics[width=\columnwidth]{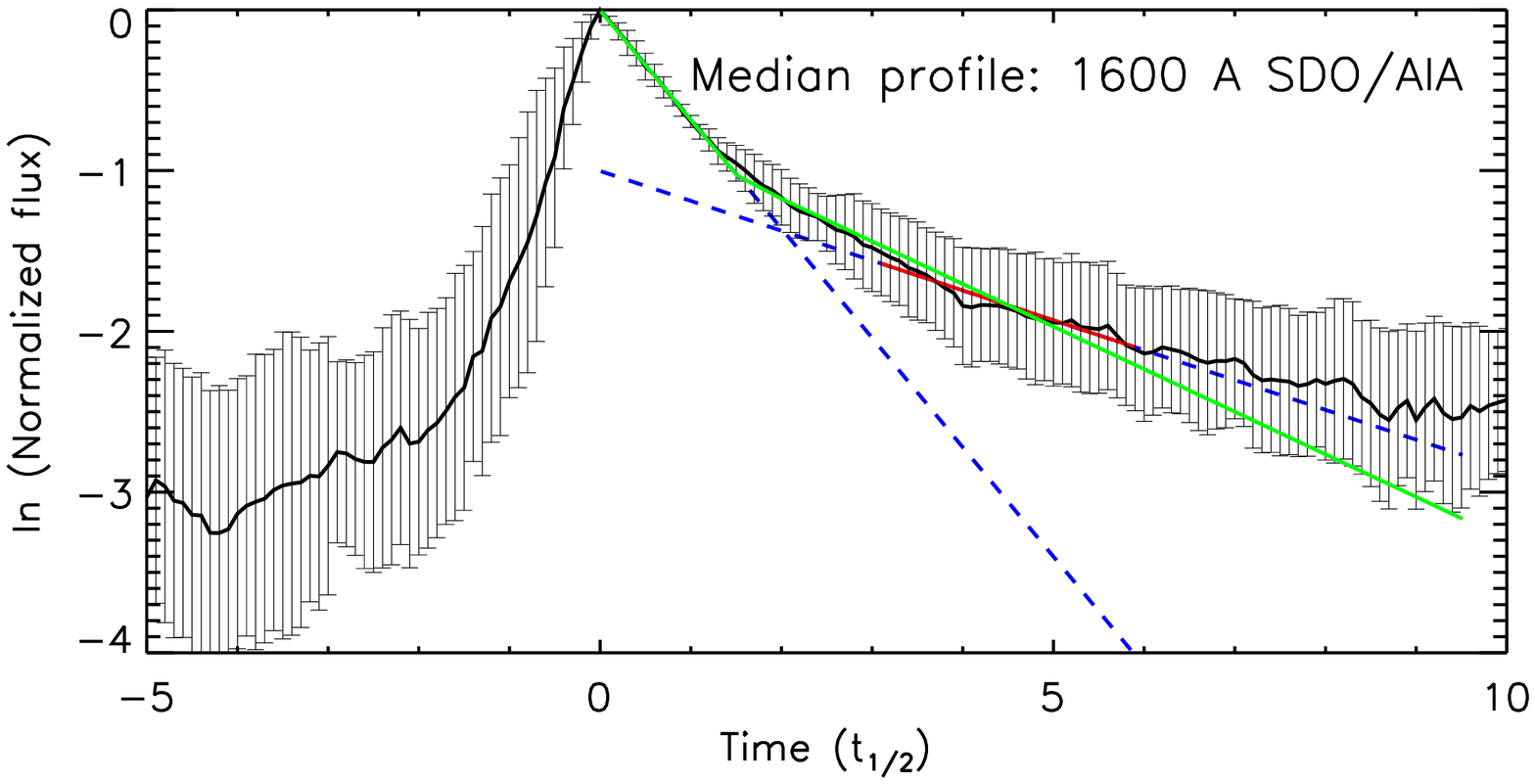}
		\includegraphics[width=\columnwidth]{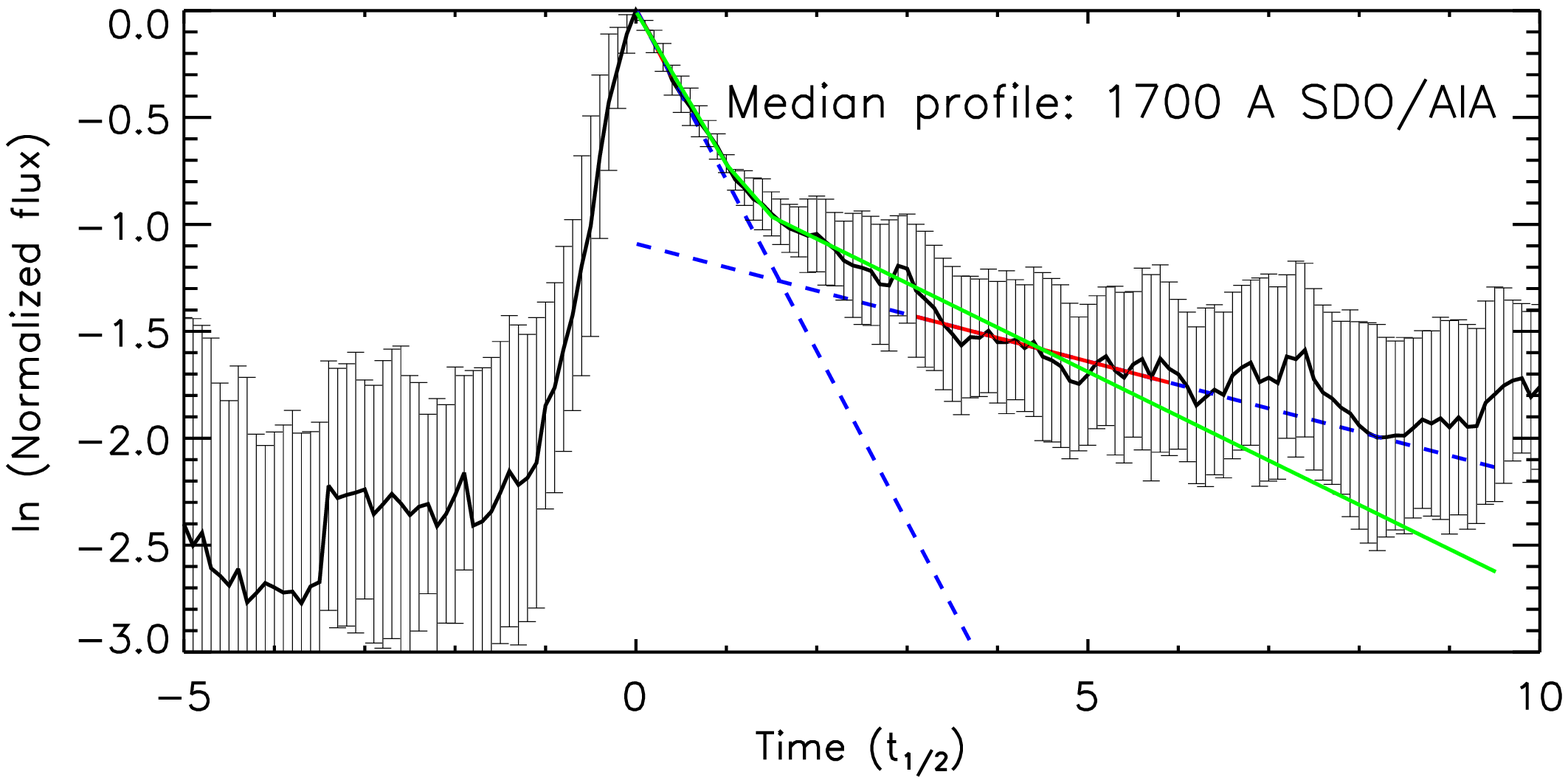}
			\includegraphics[width=\columnwidth]{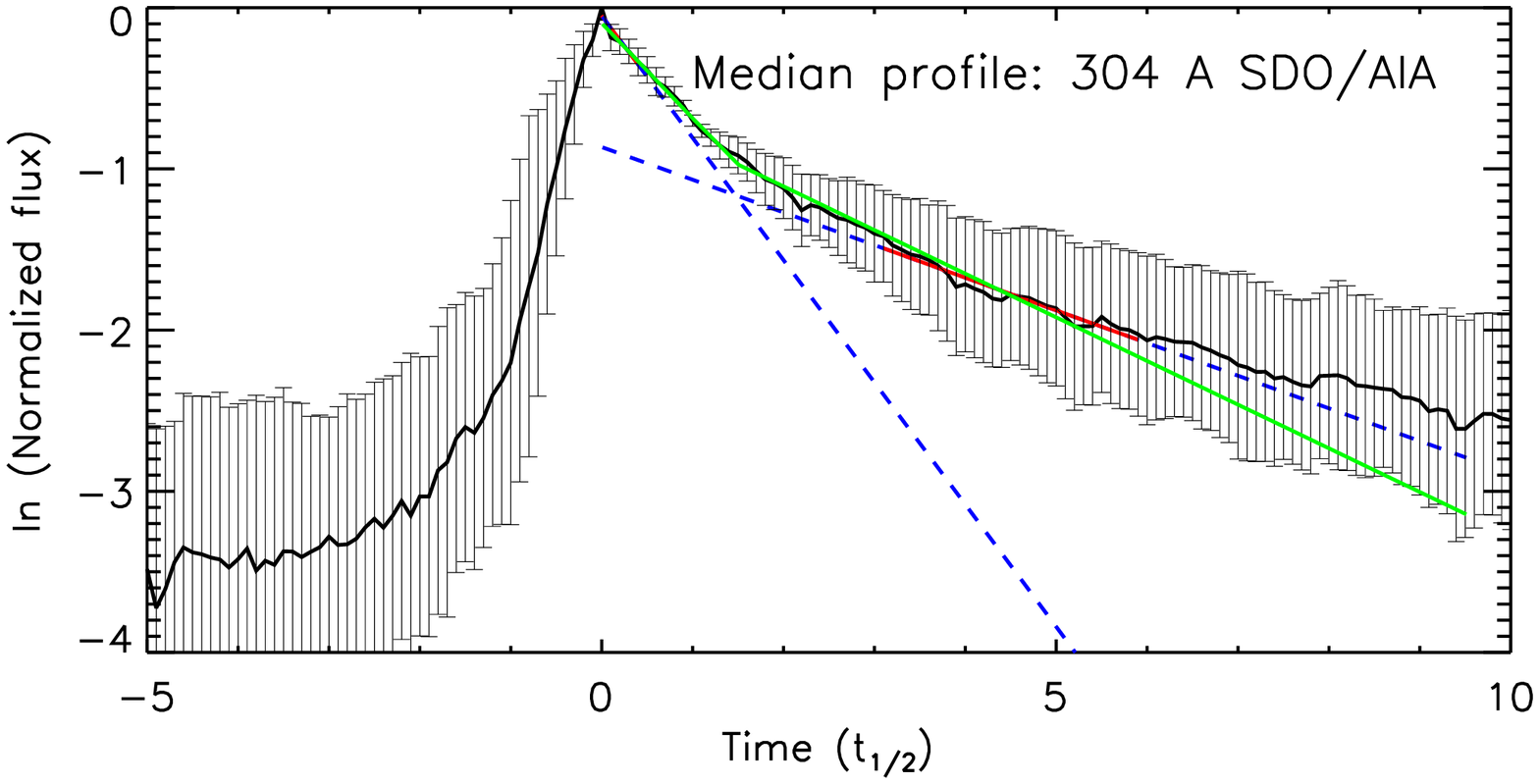}
    \caption{Average (median) time profiles (black) fitted by the broken power-law model (green line) and two exponential functions as shown in Figures \ref{fig:1600_fit}--\ref{fig:304_fit} (dashed blue lines) for t1=[0, 0.5]$t_{1/2}$ and t2=[3, 6]$t_{1/2}$.  The panels from the top to the bottom present 1600~{\AA}, 1700~{\AA} and 304~{\AA} bands, correspondingly. The red solid line marks the time interval of the second phase fitting. }
    \label{fig:1bow_fit}
\end{figure}

We also fitted the time profiles with a broken power-law model to reveal the time of the phase change and with the aim of finding a better description of the average time profiles.
The function was:
\begin{equation}
f(t)= A
   \begin{cases}
      \exp^{\textrm{\scriptsize{index$_1$}}\times t_{1/2}} & \textrm{for }t< t_{\textrm{\scriptsize{break}}},\\
      \exp^{\textrm{\scriptsize{index$_2$}}\times t_{1/2}} \times \exp^{(\textrm{\scriptsize{index$_2$}}-\textrm{\scriptsize{index$_1$}})\times t_{1/2}} & \textrm{for }t>t_{\textrm{\scriptsize{break}}}, 
     \end{cases}
     \label{eq:1bow_fit}
     \end{equation}
where $t_{\textrm{\scriptsize{break}}}$, index$_1$ and index$_2$ are all parameters determined by the fitting process.

\begin{table}
	\centering
	\caption{Results of fitting with a broken power law function }
\label{tab:table_bpow}
	\begin{tabular}{lccc} 
		\hline
		 &1600 \AA & 1700 \AA & 304 \AA \\
		\hline
	 
		index$_1$ & 0.680 $\pm$ 0.032 & 0.714 $\pm$ 0.035 & 0.589 $\pm$ 0.048 \\
	index$_2$ & 0.265 $\pm$ 0.032 & 0.207 $\pm$ 0.025  & 0.270 $\pm$ 0.034  \\
	t$_{\textrm{\scriptsize{break}}}$&	1.548 $\pm$ 0.155  & 1.282 $\pm$ 0.105 & 1.477 $\pm$ 0.198 \\
	$\chi^2$&	1.178   & 1.079 & 0.682 \\
		\hline
	\end{tabular}
\end{table}

Table~\ref{tab:table_bpow} shows the results of fitting equation \ref{eq:1bow_fit} to the median flare profiles, with $\chi^2$ values. A comparison of fitting a broken power-law model with the previous fitting, of two independent exponents, is presented in Figure~\ref{fig:1bow_fit}. One can see that the broken power-law model describing the decay phase of the average time profiles provides a good fit to the data until $t=6 t_{1/2}$. We note that the broken power-law function is a better fit to the flare profile at around 2$ t_{1/2}$.

%


	

\section{Discussion}\label{sec:discussion}
In this study we obtained averaged time profiles and fitted templates describing the decay phase of a solar flare for different spectral bands. As previously mentioned, there are two main mechanisms determining the cooling processes of a flare decay: radiative cooling and thermal conduction \citep{2004psci.book.....A}. 
According to the model of plasma cooling, thermal conduction losses initially dominates during the first part of the decay phase. Later we observe the domination of radiation losses during the second phase \citep[see,][]{Cargill1995ApJ,Aschwanden2009ApJS}. \cite{Cargill1995ApJ} also noted that if radiative losses dominate during the initial phase of decay, they still dominate during the entire decaying phase. However, here we can discriminate between the steep \citep[called impulsive by][]{Davenport2014ApJ} and gradual phases during the decay phase of the analysed time profiles. For this reason, we focused our study on the fitting of the averaged time profiles with models consisting of two components only, as done in previous studies \citep[see e.g.][]{Davenport2014ApJ}.

Domination of thermal conduction or radiative losses relates to the cooling times due to these processes, and depends on temperature, density and loop length in the case of thermal conduction. Thus, the flux behaviour during the decay phase depends on two factors~--- the ratio between cooling by radiation and cooling by thermal conduction and the ratio between the temperature and density of the generating emission region.

The temperature of spectral lines that dominate in the formation of emission of the chosen spectral bands depends on the formation height of the emission. According to \citet{ODwyer2010A&A}, emission in the 304~\AA~band is 
mainly formed by the doublet  H~II line (303.78{\AA}) that has $\log(T)=4.7$ and corresponds to the chromosphere and transition regions. We note that the SDO/AIA~{304\AA} band also has a contribution from two spectral lines with $\log(T)>6$ \cite[see][and references in it]{ODwyer2010A&A}. The first line is the Si XI 303.33~\AA~line with $\log(T) = 6.2$ whose contribution is significant in quiescent emission of off-limb active regions only. The Ca XVIII 302.19 A line with $\log(T)=6.85$ contributes to flare emission. However, its fraction of total emission was estimated as 0.05.

The quiet Sun emission in the 1600~\AA~band is generated predominantly by the C~IV line and continuum emission, both with $\log(T) = 5$. This temperature value implies that the level where the emission of this band is formed is higher than the formation height of the 304~\AA~band. The continuum dominates the emission of the 1700~{\AA} band for quiet Sun regions. Thus, the emission is formed at $\log(T) = 3.7$, and it typically originates from the photosphere. The results obtained by \citet{S2019ApJ} refined the contribution of the spectral lines emitted within the 1600~\AA~ and 1700~\AA~ bands. The authors confirmed the share of the C~IV doublet emission to the 1600~\AA~band, but they also revealed the domination of the C~I 1656~\AA~multiplet contribution over the continuum for the 1700~\AA~band. This means that emission in the 1700~\AA~ band is mostly formed at $\log(T) = 4$--4.2. Therefore, the emission from this band should be the closest in temperature to M4 dwarf flares, where $\log(T) \approx 4$ \citep{1992ApJS...78..565H}.

However, we should take into account that the structure of the atmosphere of an M dwarf star differs from the solar atmosphere. The temperature in the M dwarf atmosphere rises from cooler than the solar photosphere to hotter than the solar corona within a more narrow height range \cite[see,][]{Allred_2006ApJ}. Moreover, the initial energy release should occur lower in an M dwarf atmosphere 
than in the solar one \cite[see Figure~8a,][]{Allred_2006ApJ}.

If we fit the first phase of cooling using the same time interval 
as \cite{Davenport2014ApJ}, the coefficient $b_1$ 
of decay for solar flux is lower than that obtained by \citeauthor{Davenport2014ApJ} for the M4 dwarf even in the case of the 1700~\AA~band (0.800 \textit{vs} 0.965). Increasing the fitting interval for the first cooling phase up to 1.5$t_{1/2}$ decreased the coefficient $b_1$ 
to 0.668 for the 1700\,{\AA} band. We note that the difference between the coefficients obtained for different time intervals exceeds the error bars. Therefore, cooling during the first steep phase was found to be even slower for solar flares than for flares on the M4 dwarf. As the 1700~{\AA} channel and M dwarf flares are associated with similar temperatures this implies that the plasma responsible for M4 dwarf flare emission is denser than that associated with the 1700~{\AA} solar flares, and so potentially originates from a deeper layer. The fact that the coefficient decreases when the fitting range is extended could be related to the impact of chromospheric evaporation, which would cause an increase in flux at around 1--3$t_{1/2}$.

\begin{figure}
	\includegraphics[width=\columnwidth]{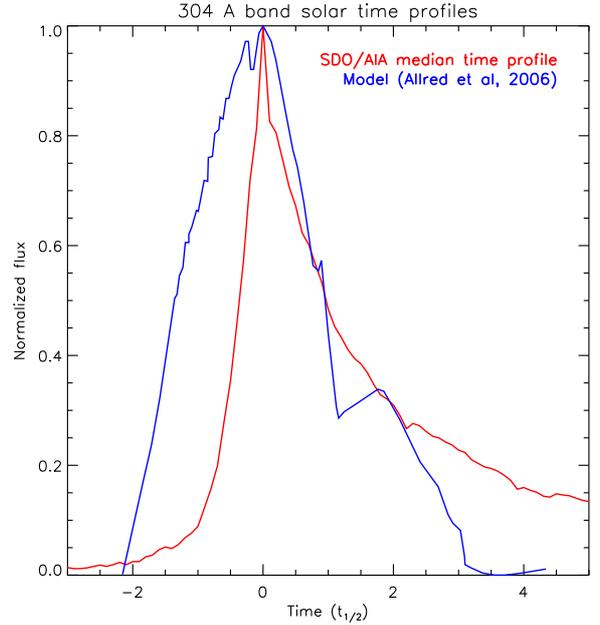}
    \caption{ Comparison of the obtained time profile of solar flare emission at 304\AA~with result of modelling by Allred et al. (2006).}
    \label{fig:304_allred}
\end{figure}

Fitting a broken power law to the 1700\,{\AA} band gave the decay coefficient index$_1$ equal to 0.714 and the time of the phase change as about 1.3$t_{1/2}$. This implies that a new phase with a different rate of cooling onsets after 1.3$t_{1/2}$. The coefficient index$_2$ for this phase, according to the fitting by a broken power law, is about 0.2. This value is close to the value obtained for M4 dwarf flare (0.29) but it does not fit time profile values above 5$t_{1/2}$. This could be due to the impact of the evaporation of the heated chromospheric plasma to the flare time profile \cite[see, for example,][]{Fl2011SSRv} .
As predicted by the modelling of \cite{Allred_2006ApJ} for the 304~{\AA} band, the contribution of the evaporation for solar flare flux is more significant than for M dwarfs. Figure \ref{fig:304_allred} contains a comparison of the 304~{\AA} band and the appropriate model from \cite{Allred_2006ApJ}, where we can see the model flux drop below the observed flux at around $t=t_{1/2}$. Then the impact of chromospheric evaporation is seen in the form of an increase in the modelled flux. Although we note the model flux again drops below the observed flux for $t>2t_{1/2}$. 


In the case of fitting for t1=[0, 0.5]$t_{1/2}$, the relation between the decay coefficients of the first phase for the 1600~\AA, 304~\AA \ and 1700~\AA \ bands confirms that the temperature of the emission formation is as discussed above. The value of $t_{\textrm{\scriptsize{break}}}$ in the broken power-law fitting characterises the transfer from the radiative cooling to the contribution of chromospheric evaporation.  We obtained the lowest value of $t_{\textrm{\scriptsize{break}}}$ for the 1700~\AA~band (1.282) and the highest value for the 1600~\AA~band. 
This is the same ordering as suggested by the temperature-height model of the solar flare atmosphere. An alternative way to specify a temperature-height model is through an analysis of the delay between maximums of flare emission in different spectral ranges. The delay between signals obtained in the UV and EUV bands are actively used for analysis of height-temperature dependence of solar atmosphere over the sunspot \citep[see,][]{Reznikova2012ApJ}. As one can see in Table \ref{tab:table0}, the delay between the maximum of flares in 304\AA~ band and 1600\AA~ band can be both positive and negative. The most common time delay was 17 seconds but a group of events demonstrated a negative delay of about 7 seconds. The uncertainties related to an image cadence of 24 seconds are 14 seconds (using the same approach as  \cite{Reznikova2012ApJ}). Thus we can conclude that negative delays are within error bars, and most of the positive delays are around this value. Such small delays, close to the level of uncertainties, agree with classical F1 and F2 flare models by \cite{Machado1980ApJ} suggesting a height difference between the layers with temperatures corresponding to 304\AA~and 1600\AA~bands of about 3 km.
 
When fitting the two exponents to the median flare  profiles, we obtained two sets of parameters for each of the first and second phases by fitting over different time intervals (see Table \ref{tab:table1}). The difference between the parameters obtained for the same phase, but for the different time intervals, exceeds the parameter uncertainty. Thus, it is necessary to determine which time intervals should be fitted to best represent the cooling processes. We believe that the decay during the second phase should be fitted using parameters obtained for t2~=~[3, 10]$t_{1/2}$ because the flux takes longer to decay in the solar atmosphere compared to the M dwarf flare profile. To reveal the decaying related to radiative and thermal conduction losses during the first phase, we should avoid the chromospheric evaporation contribution to emission. As mentioned above, the breakpoint, $t_{\textrm{\scriptsize{break}}}$ in the broken power-law fitting parameters characterises the transfer from the radiative cooling to the contribution of the chromospheric evaporation. For both the 1700 {\AA} and 304 {\AA} bands, the obtained values of $t_{\textrm{\scriptsize{break}}}$ were below 1.5. While, for these bands, the parameters obtained by fitting the different time intervals showed a significant difference, the parameters obtained by fitting the different time intervals for the 1600 {\AA} band, where $t_{\textrm{\scriptsize{break}}}$ was above 1.5, did not demonstrate a significant difference. All these facts indicate the impact of chromospheric evaporation to time profiles is not negligible (depending on observational wavelength), meaning that fitting over the interval t1~=~[0, 1.5]$t_{1/2}$ is not appropriate and the range t1~=~[0, 0.5]$t_{1/2}$ is favourable.

Based on a comparative analysis of the fitting results for solar and M dwarf flare templates, we can conclude that, for the Sun, the optimal template describing cooling processes consists of two exponents, fitted for t1~=~[0, 0.5]$t_{1/2}$  and t2~=~[3, 10]$t_{1/2}$ respectively.

We note that the second cooling phase of solar flares turned out to be more complicated than for M4 dwarf flares. The cooling of the 304~\AA~band during the second phase is marginally faster than for the 1600~\AA~band, which is hotter and originates from less dense plasma. It is possible that, during cooling the relative contribution of spectral lines with different temperatures to the emission of the 1600~\AA~band changes with time, which results in a faster decrease of emission flux. We also would like to note that the 304~\AA~band time profile demonstrated unusual behaviour. As emission of this band mostly forms by the emission of a single spectral line \citep{ODwyer2010A&A}, its time profile should be more akin to density evolution \citep[][]{Aschwanden2009ApJS}. This fact is confirmed by modelling \citep[see][]{Allred_2006ApJ}. However, the observed time profile shows an exponential decay, which is more characteristic of temperature evolution.

\begin{figure}
	\includegraphics[width=\columnwidth]{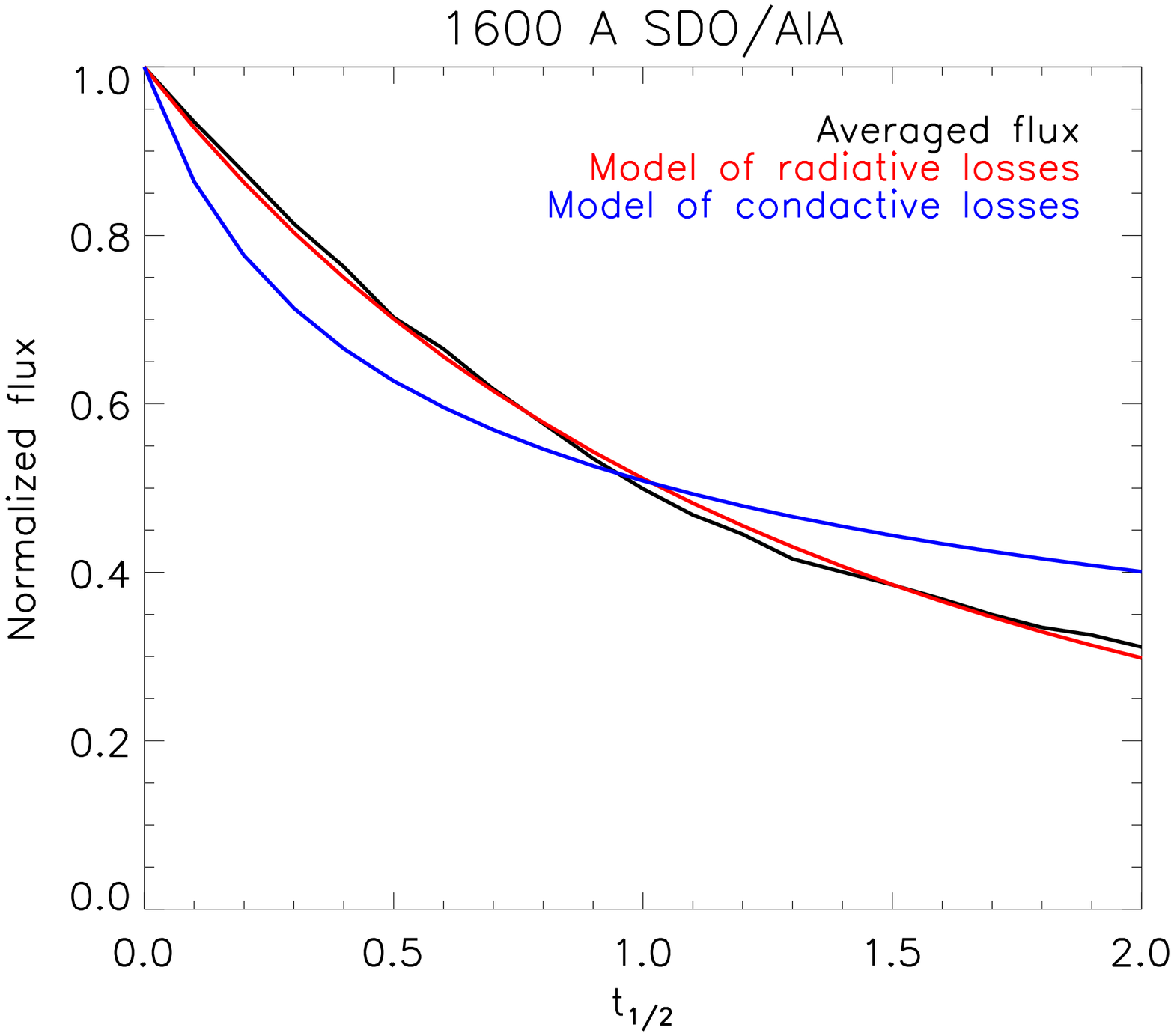}
		\includegraphics[width=\columnwidth]{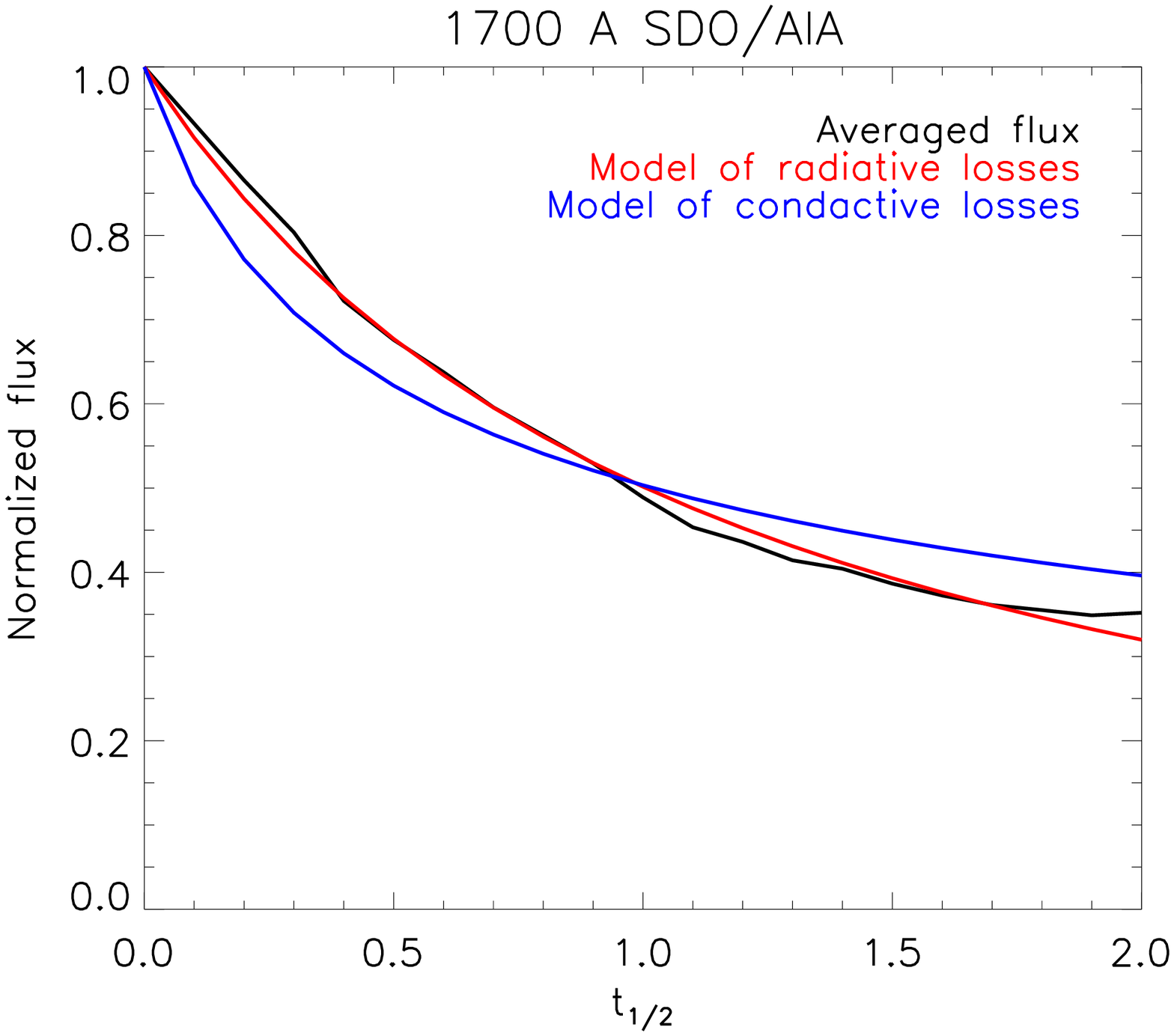}
			\includegraphics[width=\columnwidth]{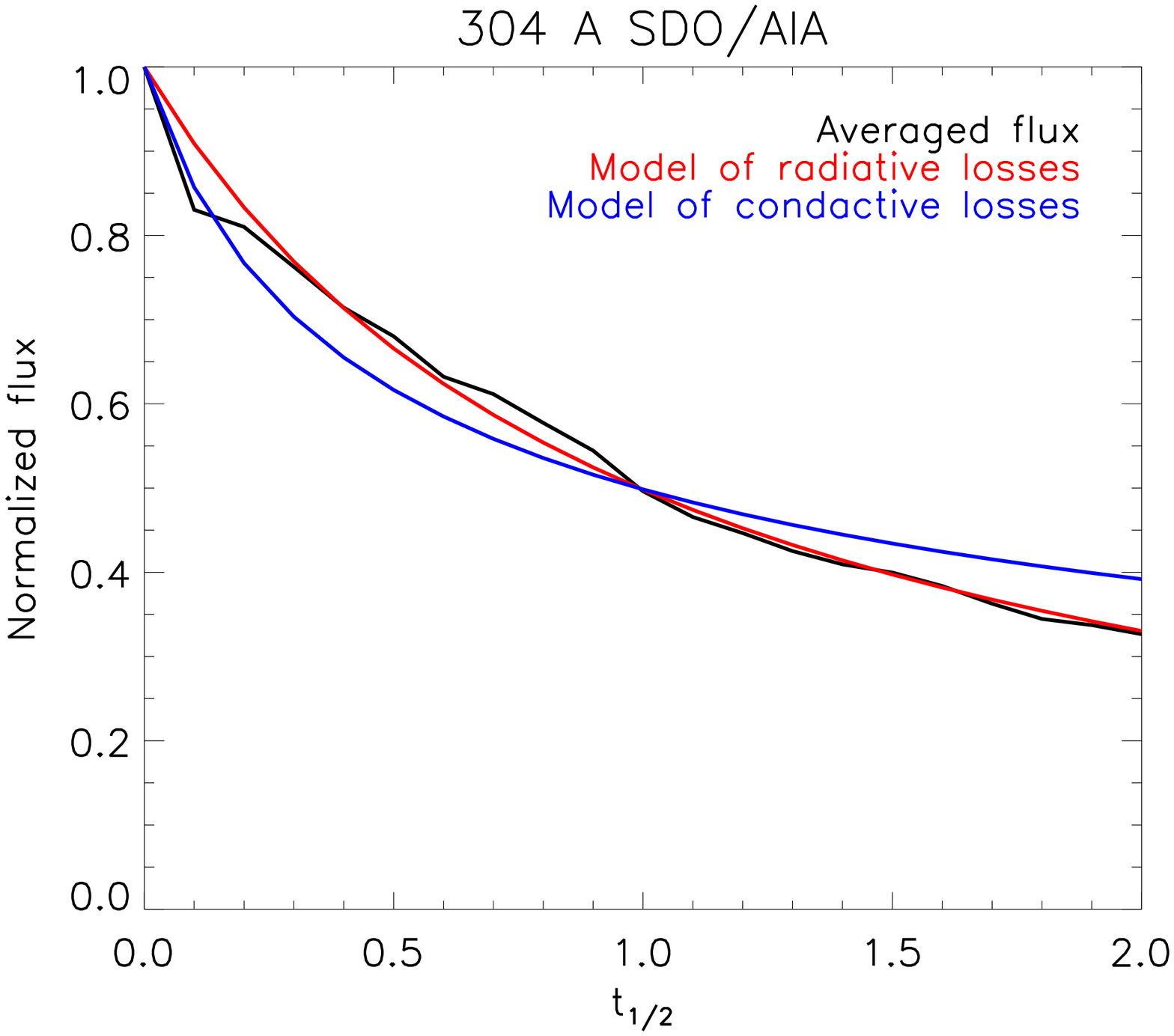}
    \caption{Average (median) time profiles (black) fitted by functions of temperature evolution depended on radiative( blue) and conductive (red) losses.}
    \label{fig:model_fit}
\end{figure}

As mentioned above, there are two processes, namely thermal conduction or radiation losses, and domination of one of the process over the other would result in different behaviours of temperature. The temperature evolution during radiative cooling can be described as  $T(s,t)=T_0(s)[1-(1-\alpha)t/\tau_{r0}]^{1/(1-\alpha)}$, where $\tau_{r0}$ is the radiative cooling time at the start of the radiative phase, $\alpha$ is the coefficient of radiative losses function and $s$ is the coordinate along the magnetic field \citep[here and after,][]{Cargill1995ApJ}. The coefficient $\alpha$ is assumed to equal to -0.5 for log(T)>5. As the temperature range of the data analysed here is less this value, we cannot use this assumption. The dependence $T(t)=T_0(1+t/\tau_{c0})^{-2/5}$ describes the temperature evolution for the case of static conductive cooling (where $\tau_{c0}$ is the conductive cooling at the beginning). We performed fitting using these two functions for the analysed spectral bands, using $\tau_{r0}$, $\tau_{c0}$ and $(1-\alpha)$ as variables. The results can be seen in Figure~\ref{fig:model_fit}. The averaged time profiles for all bands are better fitted by the function corresponding to radiative cooling. Only the 304\AA~ band time profile demonstrates good agreement with the fitting based on conductive cooling for a short time interval when $t< 0.1t_{1/2}$.

\section{Conclusions}\label{sec:conclusions}
Based on the results of reconstruction and analysis of averaged time profiles of solar flare decay phases obtained within 304~\AA, 1600~\AA~and 1700~\AA~spectral bands, we can conclude the following:
\begin{itemize}
\item The processes of cooling during the decay phase of solar flares are the most closely described by fitting of a combination of two independent exponents for t1~=~[0, 0.5]$t_{1/2}$  and t2~=~[3, 10]$t_{1/2}$. 
\item The parameters obtained for the 1700~{\AA} solar flare time profile, which is theoretically closest in temperature to the M dwarf flares, are consistent with models that imply that the white light M dwarf flare emission is formed in a higher density of layer.
\item Fitting a broken power-law model allows the contribution of chromospheric evaporation to be taken into account. However, it is not sufficient for a full description of the second part of the decay phase.
\end{itemize}

\section*{Acknowledgements}

This research was partly supported by the grant of the Russian Foundation for Basic Research No.~17-52-10001. A.M.B. acknowledges the support of the Royal Society International Exchanges grant IEC/R2/170056. This research was partly supported by the budgetary funding of Basic Research programs No. II.16 (LKK).




\bibliographystyle{mnras}
\bibliography{references1} 




\appendix

\section{Details of flares in sample} \label{sec:appendix}
Table \ref{tab:table0} contains details of the flares in our sample.

\section*{Data Availability}

The datasets were derived from sources in the public domain: SDO/AIA data was obtained from Joint Science Operations Center (JSOC) (\url{http://jsoc.stanford.edu}); GOES data were obtained from the GOES Flare Catalogue (\url{https://hesperia.gsfc.nasa.gov/goes/goes_event_listings/}).

\begin{table*}

	\caption{List of analysed events. $T_{max}$ - UT time, $t_{1/2}$ - minutes.}
		\centering
	\label{tab:table0}

 \begin{tabular}{lcccccccccc} 
 \hline
 N& Date&$T_{max}$& GOES &Location&\multicolumn{2}{c}{304 \AA}& \multicolumn{2}{c}{1600 \AA}&\multicolumn{2}{c}{1700 \AA}\\
  &yy/mm/dd&GOES& class&& $T_{max}$& $t_{1/2}$& $T_{max}$& $t_{1/2}$& $T_{max}$& $t_{1/2}$ \\
 \hline

001 & 11/02/14 & 17:26 & M2.2 & N56W18 & 17:25:23 &  4.8 & 17:25:06 &  2.4 & 17:25:20 &  4.4 \\
002 & 11/02/16 & 14:25 & M1.6 & S20W32 & 14:24:11 &  2.4 & 14:23:54 &  2.8 & 14:23:44 &  4.4 \\
003 & 11/03/08 & 22:23 & M1.0 & N31W76 & 22:20:11 &  5.0 & 22:19:54 &  4.8 &          &      \\
004 & 11/03/14 & 19:52 & M4.2 & N18W48 & 19:51:23 &  1.2 & 19:51:30 &  0.8 & 19:51:20 &  1.2 \\
005 & 11/04/15 & 13:51 & C2.8 & N13W24 & 13:50:21 &  1.0 & 13:50:18 &  0.8 &          &      \\
006 & 11/07/30 & 02:09 & M9.3 & N14E35 & 02:09:23 &  2.4 & 02:07:54 &  2.4 & 02:08:08 &  2.4 \\
007 & 11/08/04 & 03:57 & M9.3 & N19W36 & 03:54:11 & 12.6 & 03:52:42 &  6.0 & 03:54:08 &  5.6 \\
008 & 11/08/09 & 08:05 & X6.9 & N17W69 & 08:06:35 &  3.0 & 08:02:18 &  3.2 & 08:02:08 &  3.6 \\
009 & 11/09/04 & 01:07 & C8.3 & N18W79 & 01:08:11 &  3.6 & 01:05:30 &  5.2 &          &      \\
010 & 11/09/23 & 23:56 & M1.9 & N11E52 &          &      & 23:50:42 &  2.0 &          &      \\
011 & 11/09/24 & 09:40 & X1.9 & N12E60 & 09:36:21 &  1.8 & 09:36:18 &  0.8 & 09:36:08 &  1.6 \\
012 & 11/09/25 & 15:33 & M3.7 & N16E43 & 15:31:47 &  3.4 & 15:31:06 &  2.8 & 15:31:44 &  3.6 \\
013 & 11/09/26 & 05:08 & M4.0 & N13E34 & 05:07:23 &  2.2 & 05:06:42 &  2.4 & 05:06:32 &  2.4 \\
014 & 11/09/26 & 14:46 & M2.6 & N14E30 & 14:44:10 & 11.4 & 14:42:42 &  6.0 &          &      \\
015 & 11/11/15 & 22:35 & M1.1 & N20W80 & 22:32:59 &  2.0 & 22:32:42 &  3.2 &          &      \\
016 & 11/11/16 & 18:54 & C2.9 & S18E16 & 18:54:11 &  6.6 & 18:53:54 &  3.6 &          &      \\
017 & 11/11/16 & 21:43 & C1.6 & S19E13 & 21:40:45 &  3.2 & 21:40:42 &  0.8 &          &      \\
018 & 11/11/20 & 11:55 & C3.0 & S14E41 & 11:48:59 &  8.8 & 11:49:06 &  1.6 &          &      \\
019 & 11/11/22 & 04:04 & C4.9 & N13E50 & 04:02:59 &  1.4 & 04:03:06 &  0.8 &          &      \\
020 & 11/11/29 & 03:32 & C2.1 & N19E20 & 03:30:09 &  3.6 & 03:30:18 &  2.4 &          &      \\
021 & 11/12/05 & 15:18 & C4.7 & S20E00 & 15:17:47 &  1.0 & 15:17:06 &  0.8 &          &      \\
022 & 11/12/05 & 19:07 & C2.5 & S20W02 & 19:06:35 &  1.2 & 19:06:18 &  0.4 &          &      \\
023 & 11/12/18 & 02:05 & C1.9 & N19W04 & 02:05:23 &  1.8 & 02:04:18 &  2.8 &          &      \\
024 & 11/12/24 & 08:45 & C5.2 & S21W20 & 08:32:57 &  8.2 & 08:33:06 &  8.8 & 08:34:08 & 10.0 \\
025 & 11/12/25 & 18:16 & M4.0 & S22W26 & 18:15:23 &  1.0 & 18:14:42 &  1.2 & 18:14:56 &  0.8 \\
026 & 11/12/27 & 04:22 & C8.9 & S17E32 & 04:16:35 & 14.8 & 04:16:18 &  6.8 &          &      \\
027 & 11/12/28 & 14:25 & C7.2 & S23E85 & 14:23:23 &  2.2 & 14:23:30 &  3.2 &          &      \\
028 & 11/12/30 & 08:25 & C3.4 & S26E62 & 08:22:59 &  2.6 & 08:23:06 &  3.6 &          &      \\
029 & 11/12/31 & 13:15 & M2.4 & S25E46 & 13:13:23 &  2.8 & 13:13:06 &  1.6 & 13:13:20 &  2.4 \\
030 & 12/03/08 & 02:53 & C7.2 & S18W03 & 02:52:59 &  1.8 & 02:52:42 &  1.2 & 02:52:32 &  1.6 \\
031 & 12/03/14 & 15:21 & M2.8 & N14E05 & 15:19:23 & 12.4 & 15:17:06 &  6.8 & 15:16:56 &  8.8 \\
032 & 12/05/09 & 14:08 & M1.8 & N06E22 & 14:06:35 &  0.6 & 14:06:18 &  0.8 & 14:06:32 &  0.8 \\
033 & 12/05/10 & 04:18 & M5.7 & N13E22 & 04:16:58 &  0.6 & 04:16:42 &  1.2 & 04:16:32 &  2.0 \\
034 & 12/05/15 & 22:16 & C3.0 & N13W61 & 22:15:33 &  6.2 & 22:15:30 &  2.4 &          &      \\
035 & 12/06/07 & 15:43 & C9.1 & N13W06 & 15:40:35 &  7.6 & 15:37:30 &  5.6 &          &      \\
036 & 12/06/08 & 07:17 & C4.8 & N13W40 & 07:16:11 &  1.6 & 07:15:30 &  1.6 &          &      \\
037 & 12/06/29 & 06:48 & C6.2 & S16E60 & 06:47:47 &  2.0 & 06:46:42 &  3.2 &          &      \\
038 & 12/06/29 & 09:20 & M2.2 & N17E37 & 09:20:11 &  1.8 & 09:19:54 &  1.6 & 09:20:08 &  1.6 \\
039 & 12/07/01 & 19:18 & M2.8 & N14E04 & 19:15:46 &  5.4 & 19:15:53 &  4.0 &          &      \\
040 & 12/07/02 & 10:52 & M5.6 & S17E08 & 10:49:47 &  2.8 & 10:48:41 &  3.2 & 10:48:55 &  4.0 \\
041 & 12/07/05 & 11:44 & M6.1 & S22E68 & 11:43:46 &  3.4 & 11:43:53 &  1.6 & 11:43:43 &  2.8 \\
042 & 12/07/09 & 23:07 & M1.1 & S19E39 & 23:06:58 &  1.8 & 23:06:41 &  1.2 & 23:06:55 &  0.8 \\
043 & 12/07/13 & 12:19 & C1.5 & S22W27 & 12:18:58 &  1.4 & 12:17:53 &  0.4 &          &      \\
044 & 12/07/27 & 04:02 & C5.0 & N17E22 &          &      & 04:01:05 &  2.0 &          &      \\
045 & 12/10/23 & 03:17 & X1.8 & S15E59 & 03:15:46 &  0.8 & 03:15:53 &  0.4 &          &      \\
046 & 12/10/23 & 13:15 & C2.3 & S15E51 & 13:14:20 &  3.2 & 13:13:53 &  2.8 & 13:13:19 &  5.6 \\
047 & 12/11/13 & 02:04 & M6.0 & S24E47 & 02:03:22 &  2.6 & 02:02:41 &  1.2 & 02:02:31 &  1.6 \\
048 & 12/11/13 & 05:50 & M2.5 & S29E43 & 05:47:46 &  2.0 & 05:47:29 &  1.6 & 05:47:43 &  3.2 \\
049 & 12/11/17 & 18:10 & C2.8 & S24W18 & 18:09:22 &  4.4 & 18:09:29 &  3.2 & 18:11:19 &  3.6 \\
050 & 12/11/18 & 13:07 & C4.3 & N09E11 & 13:06:58 &  1.8 & 13:06:17 &  2.0 &          &      \\
051 & 12/11/20 & 12:41 & M1.7 & N11W90 & 12:38:44 &  2.0 & 12:39:29 &  0.8 &          &      \\
052 & 12/11/24 & 13:40 & C3.3 & N09W26 & 13:36:44 &  4.4 & 13:36:41 &  0.8 &          &      \\
053 & 13/01/13 & 00:50 & M1.0 & N18W18 & 00:50:10 &  0.4 & 00:48:41 &  1.6 & 00:48:31 &  1.6 \\
054 & 13/04/05 & 17:48 & M2.2 & N07E88 & 17:42:34 &  7.0 & 17:40:41 &  5.2 & 17:40:55 &  6.0 \\
055 & 13/05/02 & 05:10 & M1.1 & N10W26 & 05:04:58 &  1.8 & 05:04:41 &  1.2 &          &      \\
056 & 13/05/19 & 15:15 & C6.3 & S09W63 & 15:15:22 & 10.4 & 15:15:05 &  1.2 &          &      \\
057 & 13/05/20 & 15:06 & M1.7 & N13W08 & 15:03:46 &  2.6 & 15:03:29 &  2.4 &          &      \\
058 & 13/05/20 & 16:26 & C6.0 & N13W09 & 16:24:58 &  2.0 & 16:25:29 &  1.2 &          &      \\
059 & 13/06/24 & 11:32 & C9.9 & S17E54 & 11:31:22 &  1.6 & 11:31:29 &  2.0 & 11:31:19 &  3.2 \\
060 & 13/07/07 & 00:58 & C6.1 & S14E07 & 00:58:34 &  2.2 & 00:58:17 &  2.0 & 00:58:31 &  2.4 \\

	\hline
	\end{tabular}

\end{table*}

\begin{table*}

 \begin{tabular}{lcccccccccc} 

 \hline
 N& Date&$T_{max}$& GOES &Location&\multicolumn{2}{c}{304 \AA}& \multicolumn{2}{c}{1600 \AA}&\multicolumn{2}{c}{1700 \AA}\\
  &yy/mm/dd&GOES& class&& $T_{max}$& $t_{1/2}$& $T_{max}$& $t_{1/2}$& $T_{max}$& $t_{1/2}$ \\
 
 \hline
061 & 13/10/28 & 14:05 & M2.8 & N07W85 & 14:02:58 &  1.8 & 14:03:05 &  0.8 & 14:02:55 &  1.2 \\
062 & 13/11/02 & 04:46 & C8.2 & S23W04 & 04:44:34 &  1.8 & 04:44:17 &  1.6 &          &      \\
063 & 13/11/05 & 22:12 & X3.3 & S12E46 & 22:11:47 &  0.2 & 22:11:29 &  0.8 & 22:11:43 &  0.4 \\
064 & 13/12/29 & 07:56 & M3.1 & S18E01 & 07:53:22 &  2.2 & 07:53:29 &  1.6 & 07:53:19 &  4.0 \\
065 & 14/01/01 & 18:52 & M9.9 & S14W47 & 18:46:58 & 12.2 & 18:47:05 &  6.0 & 18:47:19 &  6.8 \\
066 & 14/01/03 & 12:50 & M1.0 & S04E52 & 12:48:58 &  5.8 & 12:49:05 &  4.4 &          &      \\
067 & 14/01/07 & 10:13 & M7.2 & S13E11 & 10:11:46 &  7.6 & 10:11:29 &  0.8 & 10:11:43 &  1.2 \\
068 & 14/01/28 & 04:09 & M1.5 & S15E88 & 04:08:10 &  1.0 & 04:07:53 &  1.2 &          &      \\
069 & 14/01/28 & 07:31 & M3.6 & S10E75 & 07:30:58 &  1.0 & 07:30:41 &  1.2 &          &      \\
070 & 14/02/01 & 01:25 & M1.0 & S11E26 & 01:23:46 &  5.4 & 01:23:29 &  3.2 &          &      \\
071 & 14/02/02 & 18:11 & M3.1 & S10E08 & 18:09:22 &  2.2 & 18:10:17 &  1.2 & 18:10:07 &  2.8 \\
072 & 14/02/07 & 04:56 & M2.0 & S15W50 & 04:54:58 &  5.4 & 04:54:41 &  4.0 &          &      \\
073 & 14/02/13 & 01:40 & M1.7 & S12W09 & 01:37:45 &  4.0 &          &      &          &      \\
074 & 14/02/13 & 06:07 & M1.4 & S11W11 & 06:06:10 &  2.8 & 06:05:53 &  3.2 &          &      \\
075 & 14/02/25 & 00:49 & X4.9 & S12E82 &          &      & 00:45:29 &  4.0 & 00:45:43 &  5.6 \\
076 & 14/03/11 & 03:50 & M3.5 & N13W55 & 03:49:46 &  3.8 & 03:49:29 &  2.0 & 03:49:19 &  2.8 \\
077 & 14/03/20 & 01:57 & C8.3 & S11E75 & 01:56:58 &  2.4 & 01:57:05 &  1.6 &          &      \\
078 & 14/03/30 & 11:55 & M2.1 & N08W43 & 11:49:46 & 13.6 & 11:48:41 &  4.4 &          &      \\
079 & 14/05/06 & 04:32 & C7.1 & S13W86 & 04:32:56 & 10.8 & 04:30:17 &  4.8 &          &      \\
080 & 14/06/11 & 08:09 & M3.0 & S14E68 & 08:06:10 &  1.8 & 08:05:53 &  2.8 & 08:05:43 &  5.6 \\
081 & 14/06/12 & 09:37 & M1.8 & S25W53 & 09:35:46 &  5.2 & 09:35:29 &  1.2 & 09:35:43 &  0.8 \\
082 & 14/08/01 & 14:48 & M2.0 & S09E35 & 14:47:47 &  0.8 & 14:47:29 &  0.4 & 14:47:19 &  0.8 \\
083 & 14/08/25 & 11:18 & B7.8 & S12E41 & 11:18:20 &  3.0 & 11:17:53 &  1.6 &          &      \\
084 & 14/10/16 & 13:03 & M4.3 & S15E84 & 13:02:34 &  0.6 & 13:02:41 &  0.8 & 13:02:31 &  1.2 \\
085 & 14/10/22 & 05:17 & M2.7 & S15E14 & 05:14:32 &  0.2 & 05:14:41 &  0.4 & 05:14:31 &  0.4 \\
086 & 14/10/23 & 09:50 & M1.1 & S16E03 & 09:47:46 &  2.6 & 09:47:29 &  0.8 & 09:47:19 &  2.0 \\
087 & 14/10/23 & 19:15 & C3.3 & S21E05 & 19:14:44 &  2.4 & 19:14:41 &  2.0 &          &      \\
088 & 14/10/30 & 17:58 & C3.4 & S05E70 & 17:56:56 &  4.0 & 17:57:05 &  3.2 &          &      \\
089 & 14/11/09 & 07:20 & C4.4 & N18E19 & 07:19:56 &  2.0 & 07:17:53 &  4.0 &          &      \\
090 & 15/01/29 & 18:15 & C5.0 & N08E63 & 18:14:20 &  2.6 & 18:14:17 &  1.2 &          &      \\
091 & 15/01/30 & 12:16 & M2.4 & N07E52 & 12:14:34 &  1.4 & 12:14:17 &  1.6 & 12:14:31 &  1.6 \\
092 & 15/03/09 & 14:33 & M4.5 & S15E49 & 14:30:58 &  1.0 & 14:31:05 &  3.2 &          &      \\
093 & 15/03/12 & 04:46 & M3.2 & S15E11 & 04:43:22 &  2.2 & 04:43:29 &  0.8 & 04:43:19 &  2.0 \\
094 & 15/05/05 & 22:11 & X2.7 & N15E79 & 22:09:22 &  1.4 & 22:08:41 &  2.0 & 22:08:31 &  2.4 \\
095 & 15/08/22 & 21:24 & M3.5 & S15E15 & 21:22:33 &  1.8 & 21:22:16 &  2.4 & 21:22:06 &  3.6 \\
096 & 15/08/28 & 19:03 & M2.1 & S13W70 & 19:03:21 &  1.0 & 19:03:04 &  1.2 & 19:02:54 &  1.6 \\
097 & 15/09/29 & 11:15 & M1.6 & S21W37 & 11:13:21 &  4.2 & 11:13:28 &  2.4 & 11:13:18 &  1.6 \\
098 & 15/09/29 & 19:24 & M1.1 & S20W36 & 19:24:09 &  2.2 & 19:23:28 &  2.0 & 19:23:18 &  2.4 \\
099 & 17/04/02 & 02:46 & C8.0 & S12W08 & 02:45:18 &  2.4 & 02:45:03 &  1.6 &          &      \\
100 & 17/04/02 & 13:00 & M2.3 & N13W61 & 12:56:56 &  3.2 & 12:56:39 &  3.2 & 12:56:29 &  6.4 \\
101 & 17/04/03 & 14:29 & M5.8 & N16W78 & 14:23:42 &  3.6 & 14:23:51 &  2.8 & 14:23:41 &  4.0 \\
102 & 17/09/05 & 01:08 & M4.2 & S09W14 & 01:06:56 &  1.4 & 01:06:39 &  2.0 & 01:06:53 &  4.0 \\
103 & 17/09/05 & 17:43 & M2.3 & S09W24 & 17:42:30 &  5.4 & 17:41:51 &  4.0 &          &      \\
104 & 17/09/07 & 10:15 & M7.3 & S07W46 & 10:15:44 &  1.0 & 10:15:27 &  1.2 & 10:15:17 &  1.6 \\
105 & 17/09/08 & 02:24 & M1.3 & S09W54 & 02:22:56 &  1.8 & 02:22:39 &  1.6 &          &      \\

		\hline
	\end{tabular}

\end{table*}



\bsp	
\label{lastpage}
\end{document}